\documentclass{ws-ijbc} 
\usepackage{graphicx}
\usepackage{epstopdf}
\usepackage{amsfonts} 					%Calls Amer. Math. Soc. (AMS) fonts

\usepackage{xcolor}
\colorlet{linkequation}{blue}
\usepackage[colorlinks]{hyperref}

\newcommand*{\refeq}[1]{%
  \begingroup
    \hypersetup{
      linkcolor=linkequation,
      linkbordercolor=linkequation,
    }%
    \ref{#1}%
  \endgroup
}
\begin{document}

\catchline{}{}{}{}{} % Publisher's Area please ignore

\markboth{Mohamed Camil Belhadjoudja}{Chaos Synchronization using Nonlinear Observers with applications to Cryptography}

\title{Chaos Synchronization using Nonlinear Observers with applications to Cryptography}

\author{Mohamed Camil Belhadjoudja}

\address{Department of Control Engineering, National Polytechnic School\\
Algiers, 16000, Algeria\\
mohamed\_camil.belhadjoudja@g.enp.edu.dz}

\maketitle

\begin{history}
\received{(to be inserted by publisher)}
\end{history}

\begin{abstract}
The goal of this survey paper is to provide an introduction to chaos synchronization using nonlinear observers and its applications in cryptography. I start with an overview of cryptography. Then, I recall the basics of chaos theory and how to use chaotic systems for cryptography, with an introduction to the problem of chaos synchronization. Then, I present the theory of non-linear observers, which is used for the synchronization of chaotic systems. I start with an explanation of the observability problem. Then, I introduce some of the classical observers: Kalman filter, Luenberger observer, Extended Kalman filter, Thau's observer, and High gain observer. I finish by introducing the more advanced observers: Adaptive observers, Unknown inputs observers, Sliding mode observers and ANFIS (Adaptive Neuro-Fuzzy Inference Systems) observers. 
\end{abstract}


\section{Introduction}
The subject of this paper is the synchronization of chaotic systems based on non-linear observers for applications in cryptography. 
\\ \\ Cryptography is the science of secure data transmissions. It is part, along with cryptanalysis -which is the science that aims to test the security of cryptographic systems (cryptosystems)- of the broad field of cryptology. 
\\ \\ We essentially find two categories of techniques for the design of cryptosystems: mathematical techniques and physical techniques. The vast majority of mathematical techniques are related to number theory, elliptical curves, abstract algebra, and more recently the theory of lattices. The physical techniques are focused on the use of notions of quantum mechanics such as the Heisenberg principle and the polarization of photons.
\\ \\ Over the years, we have seen the emergence of a new mathematical technique for the design of cryptosystems, namely chaos theory. This emergence gave birth to what is today called chaotic cryptography, which is still at the experimental stage and which could in the coming years be a significant support to already existing encryption systems. 
\\ \\ A chaotic system can be implemented in two different ways. In the form of a computer program (C code for example) or in physical form (an electrical circuit). The problem with the computer implementation is that there is what is called the phenomenon of dynamic degradation. Because finite precision is used in a computer implementation, the program may behave differently from the actual chaotic system. For this reason, it is preferable to use a physical realization of the chaotic system based on electronic components. \\ \\ In cryptography, there is an emitter that transmits the encrypted message and a receiver that decrypts it. In chaotic cryptography, to perform these encryption and decryption operations, we need to reproduce the same chaotic signal at the emitter and the receiver. A first idea would be to design the same electrical circuit with the same parameters. The problem with this idea is that it is impossible to reproduce with infinite precision the same circuits, there will always be uncertainty and noise which will cause differences between the parameters of the two chaotic circuits. And because of the strong dependence on the initial conditions of chaotic systems, the slightest difference, however small it may be, between the parameters of the two chaotic circuits will cause a large difference between the chaotic signals created at the level of the emitter and the level of the receiver. To overcome this, we use chaos synchronization. 
\\ \\ Chaos synchronization consists of ensuring that a dynamical system called the slave system reproduces with a certain precision the signals emitted by a chaotic circuit called the master system. There are several chaos synchronization techniques. among these techniques, we find those based on nonlinear observers.
\\ \\ An observer is a dynamical system whose role is to reproduce certain signals based on partial information on the system which initially emitted these signals. The idea of synchronization based on non-linear observers is to take as the slave system an observer whose objective is to reproduce the signals emitted by a chaotic circuit.
\section{Cryptography}
Cryptography is the science behind the design of secure communication systems. We find it for example in the military, political, industrial, and medical fields where there is a transmission of private data. The main goal of cryptography is to transform the data into a code that is not understandable by an eventual attacker. For example, in World War 2, the Nazis used a machine called ENIGMA for secure military communications. The problem with ENIGMA is that it had some weaknesses (for example, a letter can be anything in the encrypted version except itself), which allowed the allies to break it. According to specialists, breaking ENIGMA shortened the war by at least 2 years. Cryptography is, therefore, an area of capital importance that it is essential to continuously develop. 
\subsection{Cryptosystems} 
\begin{figure}
  \centering
  % include first image
  \includegraphics[width=1\linewidth]{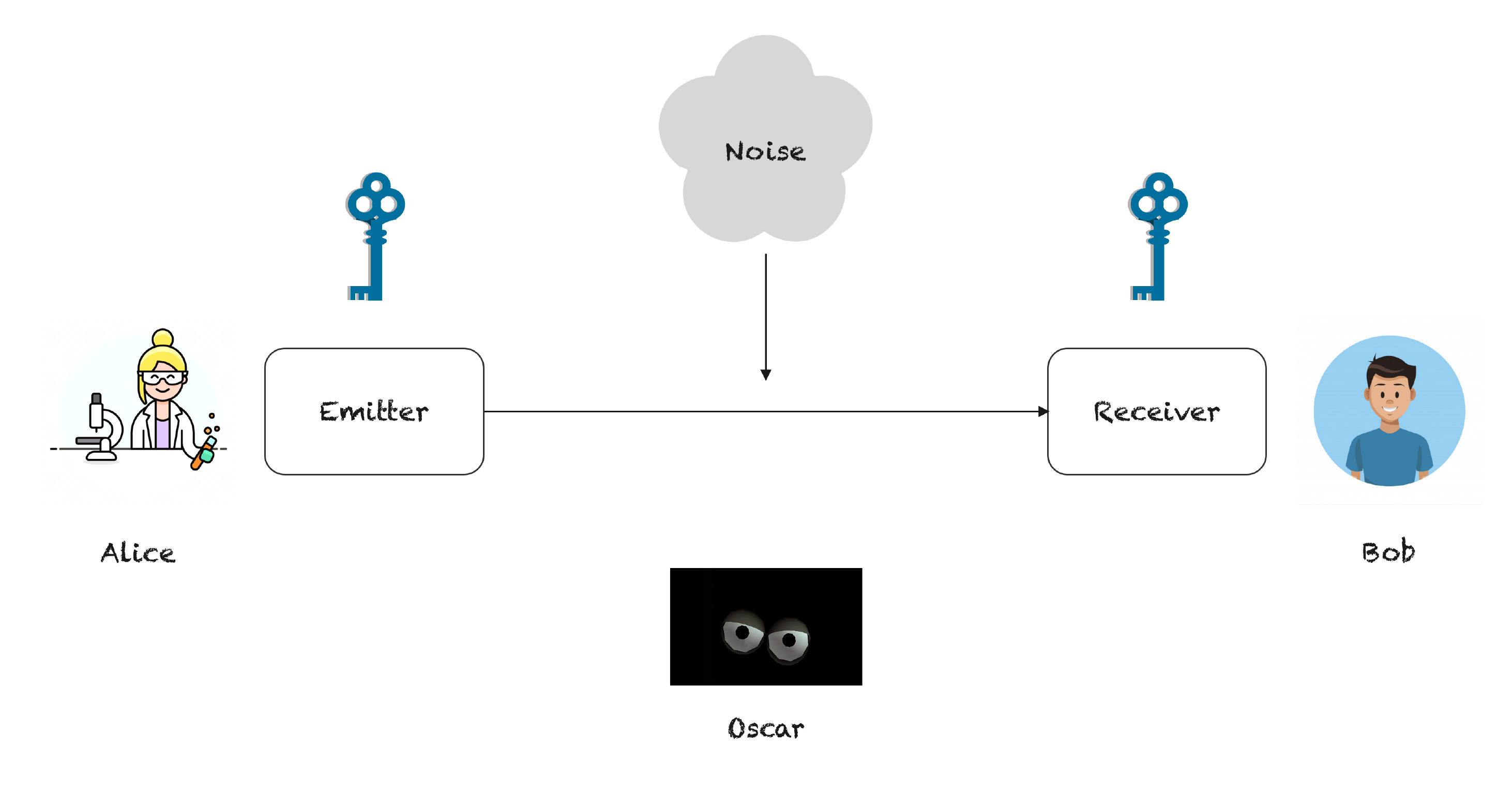}
  \caption{The structure of a cryptosystem}
  \label{The structure of a Cryptosystem}
\end{figure}

\begin{figure}
    \centering
    \includegraphics[width=0.7\linewidth]{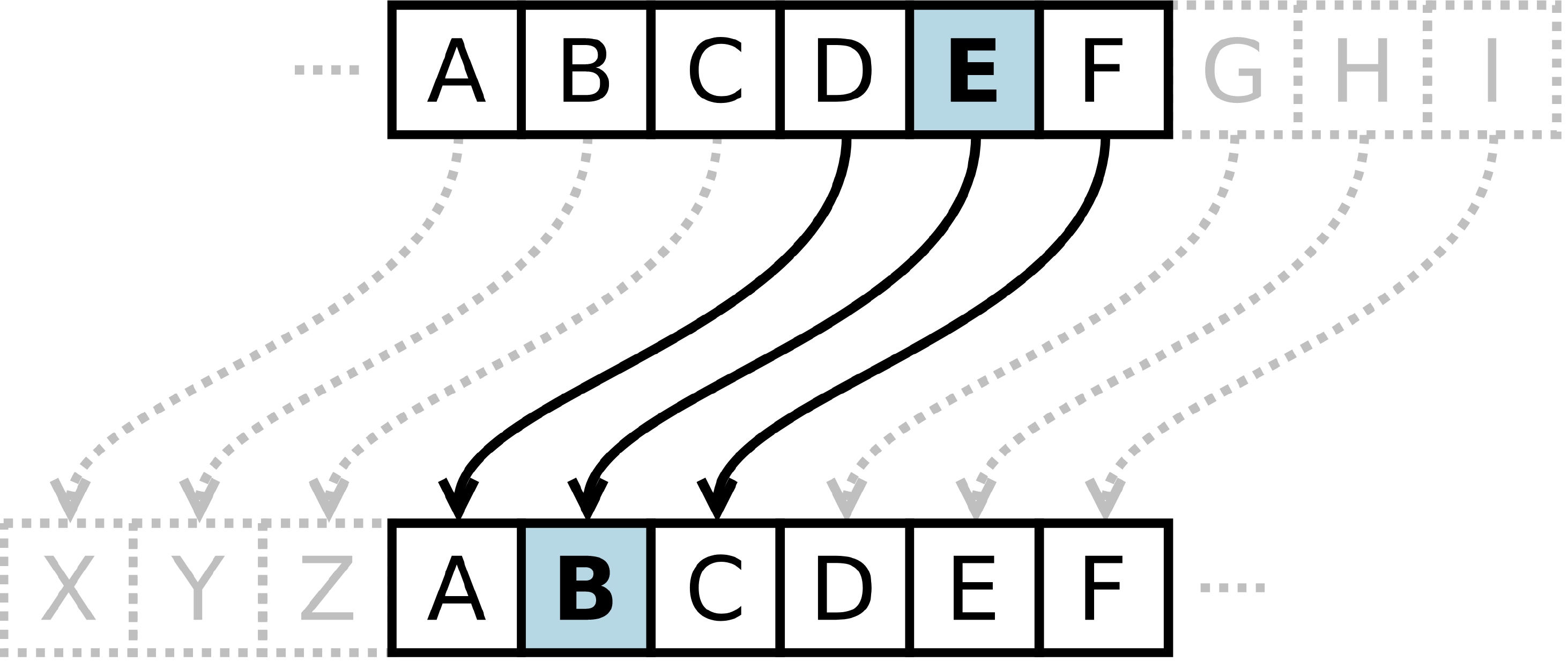}
    \caption{The Shift Cipher}
    \label{The shift cipher}
\end{figure}

\begin{figure}
    \centering
    \includegraphics[width=0.7\linewidth]{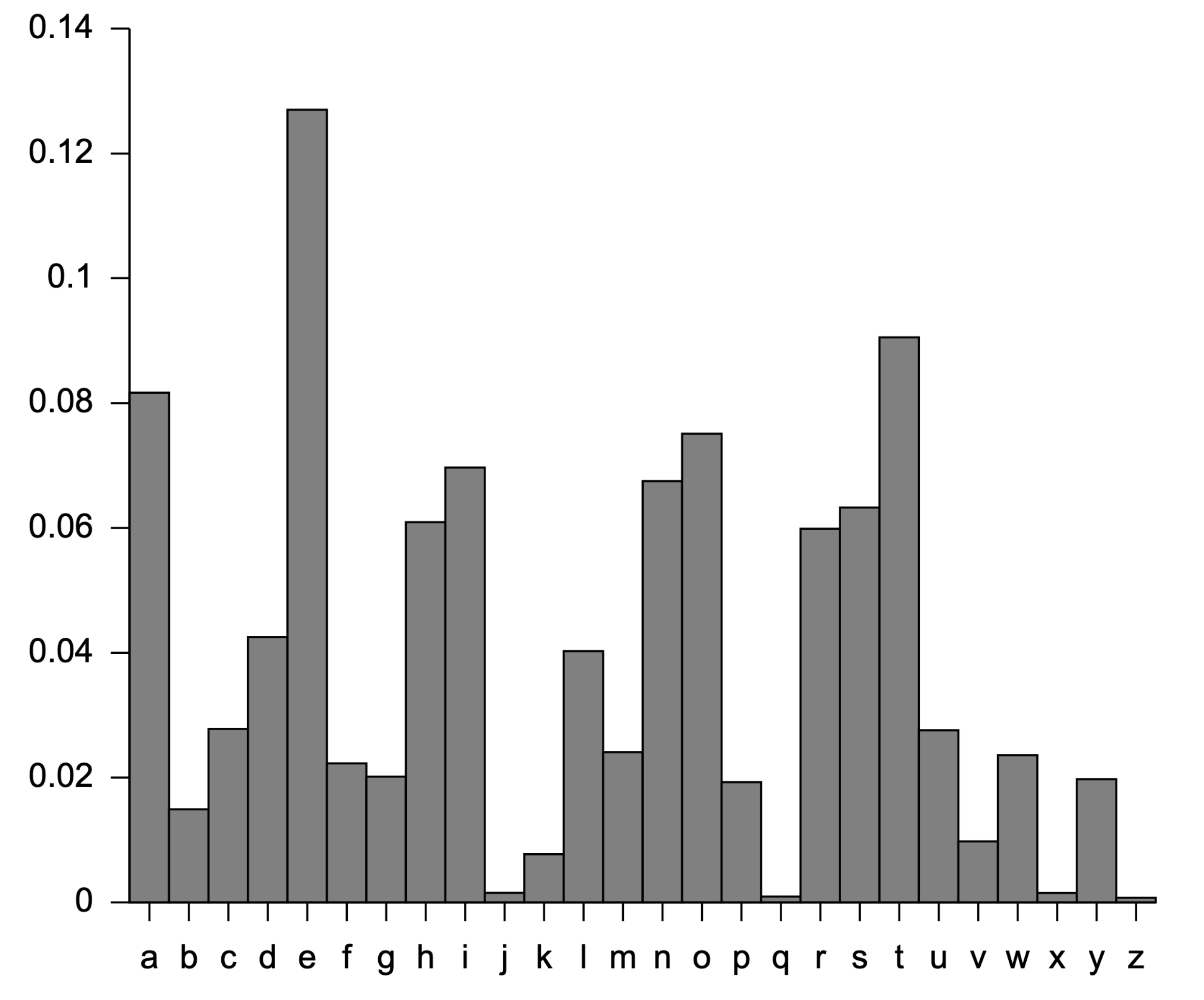}
    \caption{The frequency of appearance of letters in English}
    \label{The frequency of apparition of letters in english}
\end{figure}

The fundamental element in cryptography is the cryptosystem whose structure is shown in figure \ref{The structure of a Cryptosystem}. Alice wants to send a message (the plaintext) to Bob but does not want this message to be read by a spy who will be called Oscar. To send her message, Alice uses an emitter which transforms the plaintext into a code (a ciphertext) and Bob uses a receiver to decode the ciphertext and read the message. To operate the emitter and receiver, Alice and Bob must use a key. The receiver's key is kept secret, in the sense that only Bob knows it. The emitter's key can be secret, and in this case, it is generally the same as Bob's and we talk about symmetric cryptography, or public, and in this case, we talk about asymmetric cryptography or public-key cryptography. 
\\ \\ I will describe now a historical cryptosystem: The Shift Cipher, whose principle is shown in figure \ref{The shift cipher}. The Shift Cipher is the cryptosystem that was used in ancient Rome by Caesar when he wanted to communicate with his generals. First, each letter is associated with a number, and the plaintext is thus translated into a series of numbers. Then, we choose a key between 0 and 25, which must only be known by the two sides of the communication system. After that, each number of the plaintext is added modulo 26 to the key. The sequence obtained is translated into letters, which give us the ciphertext that can be sent. To decrypt the code, we transform the ciphertext into a series of numbers, and we subtract modulo 26 from each number the key, then we translate the result into a series of letters. The result is the plaintext. In the Shift Cipher, the emitter and receiver have the same key and it must be kept secret otherwise anyone will be able to easily understand the ciphertext.
\\ \\ I will introduce now cryptanalysis, the science that studies the security of cryptosystems. The fundamental idea when we design a cryptosystem is The Kerckhoffs principle. According to this principle, to have a secure communication system, it is necessary when we design it to assume that the attacker will know the encryption and decryption algorithm, except the secret key, and that he will have access to the emitter and the receiver. In other words, we should minimize the number of secret components in a cryptosystem, and we must not suppose that the architecture of the cryptosystem is secret\footnote{In general it is easy to find it using techniques like reverse engineering.}.
\\ \\ To talk about the security of a cryptosystem, we must study the key space, plaintext space and ciphertext space. These terms are explained in the following definition: \\ \\
\textbf{Definition 1 \cite{paar}:} \textit{ Cryptosystem \\
A cryptosystem is a five-tuple of sets ($\mathcal{P}$,$\mathcal{C}$,$\mathcal{K}$,$\mathcal{E}$,$\mathcal{D}$) such that : \\ 
1- $\mathcal{P}$ is a finite set of possible \textbf{plaintexts}. \\ 
2- $\mathcal{C}$ is a finite set of possible \textbf{ciphertexts}. \\ 
3- $\mathcal{K}$, the \textbf{keyspace}, is a finite set of possible \textbf{keys}. \\ 
4- For each $K\in \mathcal{K}$, there is an \textbf{encryption rule} $e_{K}\in \mathcal{E}$ and a corresponding \textbf{decryption rule} $d_{K}\in \mathcal{D}$. Each $e_{k} : \mathcal{P} \to \mathcal{C} $ and $d_{K} : \mathcal{C} \to \mathcal{P}$ are functions such that $d_{K}(e_{K}(x)) = x$ for every plaintext element $x\in \mathcal{P}$. }
\\ \\ Let us now analyze the security of the shift cipher. The first method to break the shift cipher is the Brute force attack. We test several keys until we find the right one. In practice, in secure cryptosystems the keyspace is so large that it is almost impossible to find the right key in this way. However, in the case of the shift cipher, there are only 26 keys to test, which makes the brute force attack relatively effective. \\
The second method of cryptanalysis is frequency analysis. A frequency diagram of the appearance of letters in English is shown in figure \ref{The frequency of apparition of letters in english}. It can be used, with a diagram of the appearance of terms in English, to compare the frequency of appearance of letters and terms in the ciphertext with the frequency of appearance of letters and terms in English (or in general in the language used for communication). For example, the letter most used in English is E, we can then assume that the letter that is most used in the ciphertext corresponds to the encrypted version of E (as long as the ciphertext is sufficiently large). 
\\ Another useful method for basic cryptosystems such as the shift cipher is to choose a plaintext, get the associated ciphertext, change a single letter of the plaintext, and see how the ciphertext varies. For the shift cipher, by changing a single letter of the plaintext it will change a single letter of the ciphertext, we can therefore deduce that this letter is the encrypted version of the letter that we changed in the plaintext. This method, just as the frequency analysis attack, uses the weaknesses of the internal structure of the algorithm. It is classified as a mathematical analysis attack. There are many other attacks for different cryptosystems. We can use for example implementation attacks, which are techniques using the weaknesses of the hardware and software that are used. Another category of attacks is social engineering where we use the vulnerabilities of the human mind.
\subsection{Symmetric cryptography}
\begin{figure}
    \centering
    \includegraphics[width=1\linewidth]{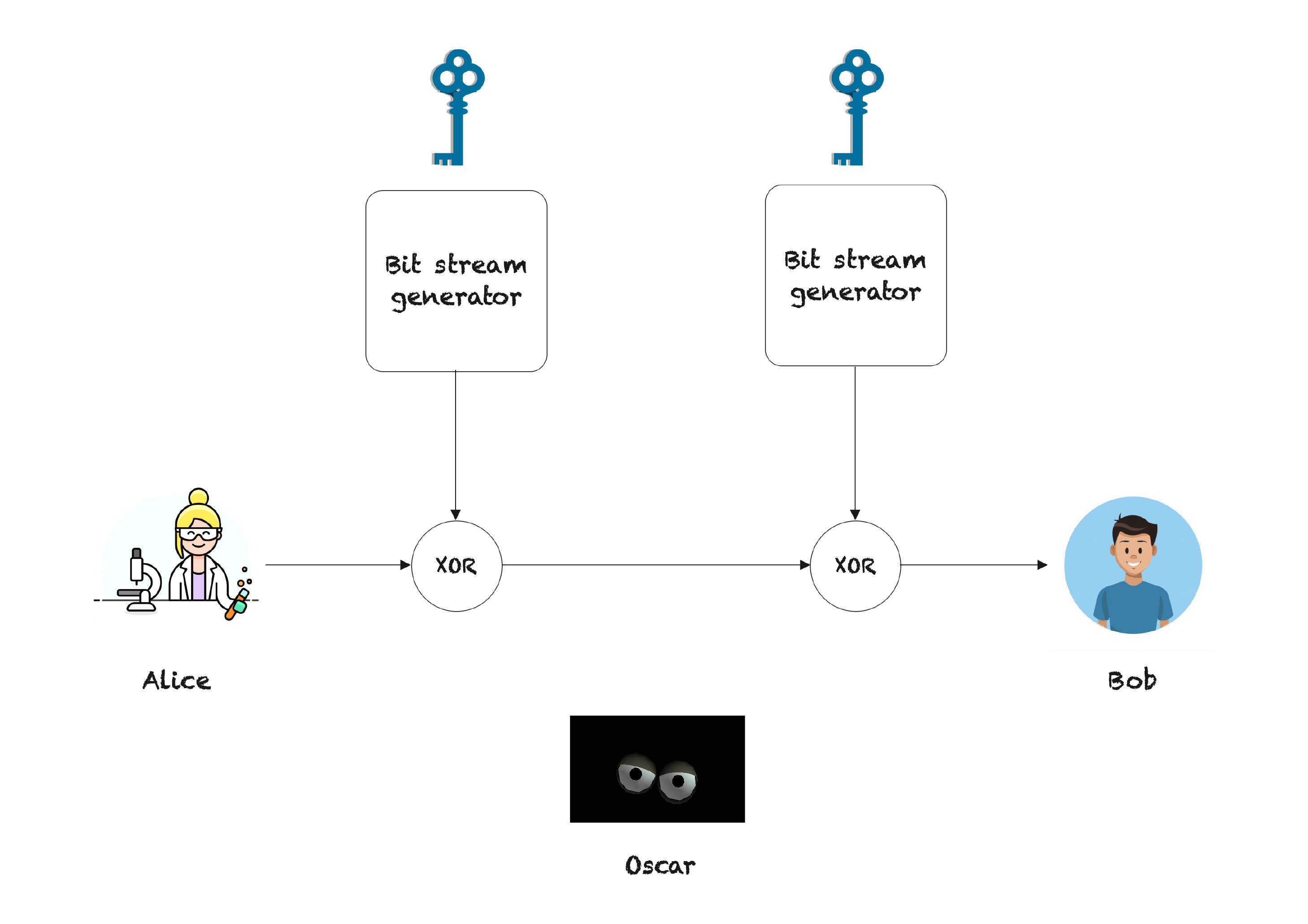}
    \caption{The structure of a stream cipher}
    \label{streamcipher}
\end{figure}

\begin{figure}
    \centering
    \includegraphics[width=1\linewidth]{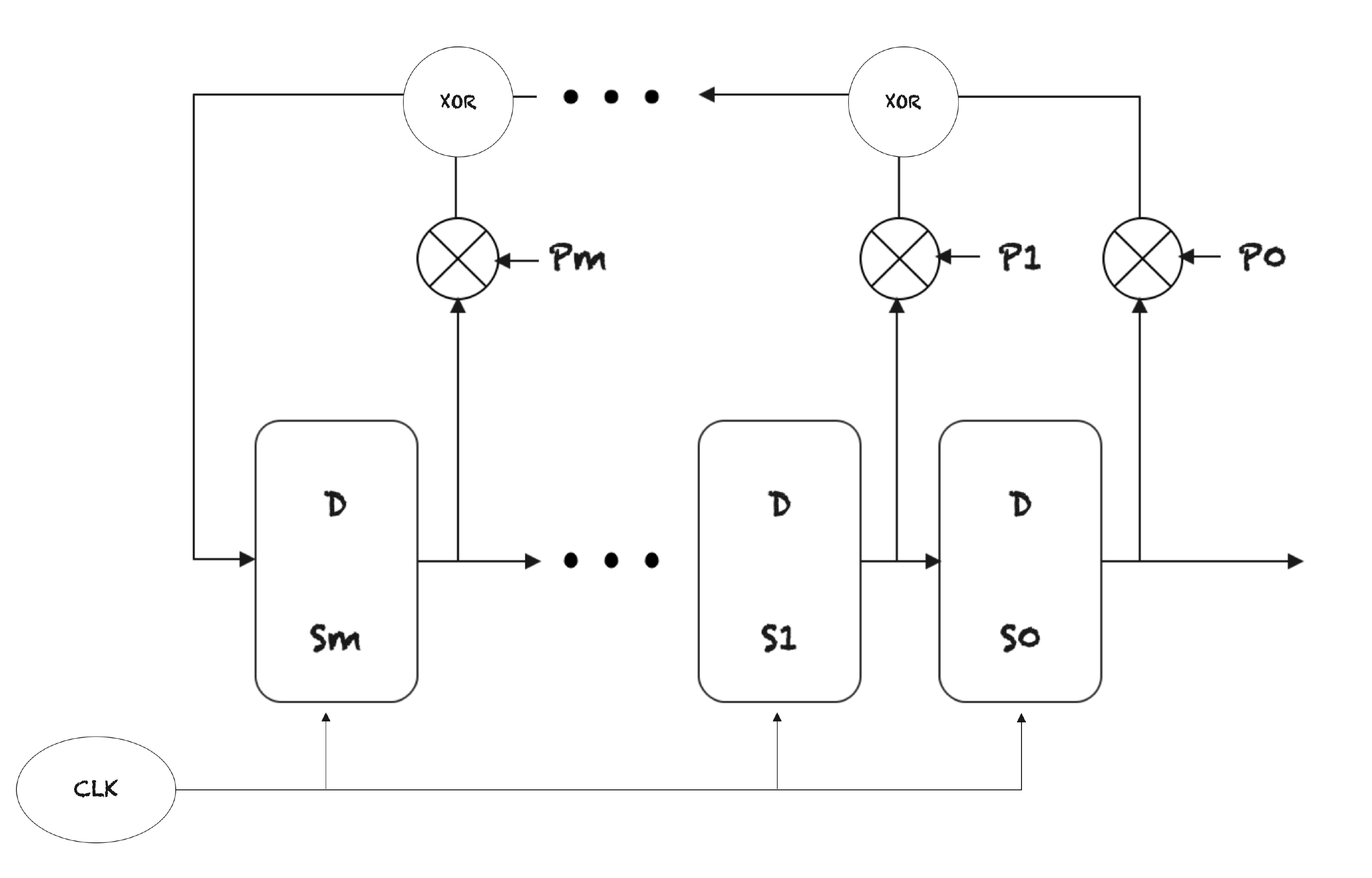}
    \caption{The structure of a LFSR}
    \label{lfsr}
\end{figure}

There are two types of symmetric cryptosystems. First, we have the Stream ciphers. The structure of a stream cipher is shown in figure \ref{streamcipher}. Stream ciphers are cryptosystems where we encrypt one bit each time and send it to the receiver for decryption. This operation is realized using XOR gates and bitstream generators. We XOR each bit of the plaintext with the bit generated by the bitstream generator, and then we do the same thing at the receiver side to recover the plaintext bit. The secret key is the input of the bitstream generator, and because we must have the same bitstream added at the receiver and the emitter side, we need the same key on the two sides, it is the reason why this cryptosystem is symmetric.
\\ \\ The bitstream generator needs to be random (TRNG: True Random Number Generator), or at least pseudo-random (PRNG: Pseudo-Random Number Generator). It also must have a fundamental property that ensures the security of this algorithm: unpredictability, which means that even if we know of any number of bits generated by the bitstream generator, it is computationally infeasible to compute the next (or preceding) bits from this information. These properties are here to make it impossible to guess the bits and decrypt the message without the key. An unpredictable PRNG is also called a CSPRNG (Cryptographically Secure Pseudo-Random Number Generator) \cite{paar}. The security of a stream cipher will depend mostly on the statistical properties of the CSPRNG used. We should ensure that the bitstream generator has good randomness properties and that it is unpredictable. There are mathematical tests of randomness like the Diehard tests and the chi-square test. There are also software like TestU01 that are dedicated to randomness tests. For unpredictability, we must ensure that the relationship between the bits generated by the bitstream generator is sufficiently complex.
\\ \\ The PRNG can be implemented into a CPU or it can be realized uniquely with hardware components. An example of a PRNG that is implemented in hardware is the Linear Feedback Shift Register (LFSR), whose structure is shown in figure \ref{lfsr}. 
 \\ \\ In a LFSR, there are flip-flops, XOR gates, and multiplication symbols. The multiplication symbols are here to act as switches. If $p_{i}=0$ then the switch is open, and if $p_{i}=1$ the switch is closed. The bitstream is the sequence of outputs of the last flip-flop ($s_{0}$). The mathematical description of the input of the leftmost flip-flop is given by \refeq{lfsrequation}. 
\begin{equation} \label{lfsrequation}
s_{m+1} \equiv s_{m}p_{m} + ... + s_{1}p_{1} + s_{0}p_{0} \ mod \ 2
\end{equation}
The maximum sequence length generated by this type of PRNG is $2^{m+1} -1$, where $m+1$ is the number of flip-flops. The maximum length is however generated only by some combinations of $p_{i}$s. For example, for $m+1=4$, $p_{3} = p_{2} = 0 $ and $ p_{1} = p_{0} = 1$, the PRNG has a sequence of maximum length, namely $2^4 - 1 = 15$. \\
The vector $(p_{0}, p_{1}, ..., p_{m-1},p_{m})$ is the secret key. It is common to represent it as a polynomial with coefficients in $\lbrace 0,1 \rbrace $ as in \refeq{poly}. 
\begin{equation} \label{poly}
P(x) = x^{m+1} + p_{m}x^m + p_{m-1}x^{m-1} + ... + p_{1}x + p_{0} 
\end{equation} 
We can show that maximum-length PRNGs have primitive polynomials which can easily be computed. Thus, it is easy to find the maximum-length PRNGs.
\\ \\ It is clear that if we have $m+1$ different equations of the form \refeq{lfsrequation}, it will be easy to compute the secret key (because it will become a system of $m+1$ linear equations with $m+1$ unknowns). It is one of the weaknesses of LFSR-based stream ciphers. The bitstream generator is not unpredictable. One way of overcoming this is to make the $p_{i}$s random (generating continuously random keys).
\\ \\ The other symmetric cryptosystems used in practice are Block ciphers. Block ciphers are the most secure and popular symmetric ciphers in cryptography. Unlike a stream cipher where we proceed one bit at a time, with block ciphers we encrypt a group of bits at once. The most popular block ciphers are AES (Advanced Encryption Standard), DES (Data Encryption Standard) and their modifications (3DES, ...). For more details about block ciphers, you can refer to \cite{paar}. I will not detail the structure of block ciphers in this paper. I have mentioned them only to introduce the concepts of confusion and diffusion which are fundamental in cryptography \cite{Shannon}:
\\ \\ - \textbf{Confusion} : It is an encryption operation where the relation between the key and the ciphertext is obscured. \\
- \textbf{Diffusion} : It is also known as the avalanche effect. If we change one bit of plaintext, it must affect many ciphertext bits. It is done to mask the statistical properties of the plaintext. 
\subsection{Public key cryptography}
Public key cryptography is a branch of cryptography where the emitter's key is public and the receiver's key is private. We call such cryptosystems asymmetric cryptosystems or public-key cryptosystems. Their main interest is that anyone can send a message, but only one person can decrypt it. It is for example used during banking transactions where anyone can send information to a bank but only the bank is able to read this information. For details about the public key cryptosystems, you can refer to \cite{paar}. I will explain what the security of the most popular public-key algorithms is based on.
\\ \\ The most popular and possibly the most secure public-key cryptosystem is RSA (Rivest-Shamir-Adleman). The security of RSA is essentially based on the problem of factorization of integers which is stated as follows: Given an integer $n$, find the two primes $p$ and $q$ such that $n = pq$. It is considered to be a difficult problem, in the sense that no current computer could solve this problem in a suitable time. For current computers, the best published algorithm for integer factorization is \textbf{GNFS} (general number field sieve) that runs on a $b$-bit number $n$ in time: $exp(((64/9)^{1/3}+o(1))(ln \ n)^{1/3}(ln \ ln \ n)^{2/3})$. 
\\ \\ Another public-key algorithm is the Diffie-Hellman key exchange. It is a public key algorithm that allows Alice and Bob to create the same secret key that they can use in a symmetric encryption algorithm. The security of the Diffie-Hellman key exchange is based on the discrete logarithm problem which is stated as follows: Given an integer $t$ and a generator $g$ of $\mathbf{Z}/m\mathbf{Z}$, compute $l = log_{g}t$. The discrete logarithm problem is considered, like the factorization of integers, to be a difficult problem.
\\ \\ With the advent of quantum computers, public-key cryptosystems are in danger. As an example, Shor \cite{Shor} showed that using a quantum computer, it is possible to solve the factorization problem and the discrete logarithm problem in a reasonable time. This gave rise to quantum cryptography, which is based on quantum mechanics, and post-quantum cryptography, which is based on mathematical methods resistant to quantum technology, such as certain optimization problems on lattices. In addition to the problems associated with public-key encryption algorithms, there are cryptanalysis methods that exploit weaknesses in the Hardware. For example, for RSA, we find in \cite{Acoustic} an acoustic cryptanalysis technique that exploits the sound emitted by RSA decryption in order to reconstruct the secret key. There are also cryptanalysis techniques that exploit the energy consumed by certain operations of RSA algorithm, such as the rapid exponentiation algorithm, to break the code. These techniques are called Power Analysis techniques, or Side Channel attacks. 
\section{Chaotic cryptography} 
In this section, I present basic notions on chaos theory and its use in cryptography while highlighting the problem of chaos synchronization. I start by qualitatively defining what a chaotic system is and what it means to synchronize chaos, then I explain how chaos is used in cryptography in current research.
\subsection{Chaotic Systems}
\begin{figure}
  \centering
  % include first image
  \includegraphics[width=1\linewidth]{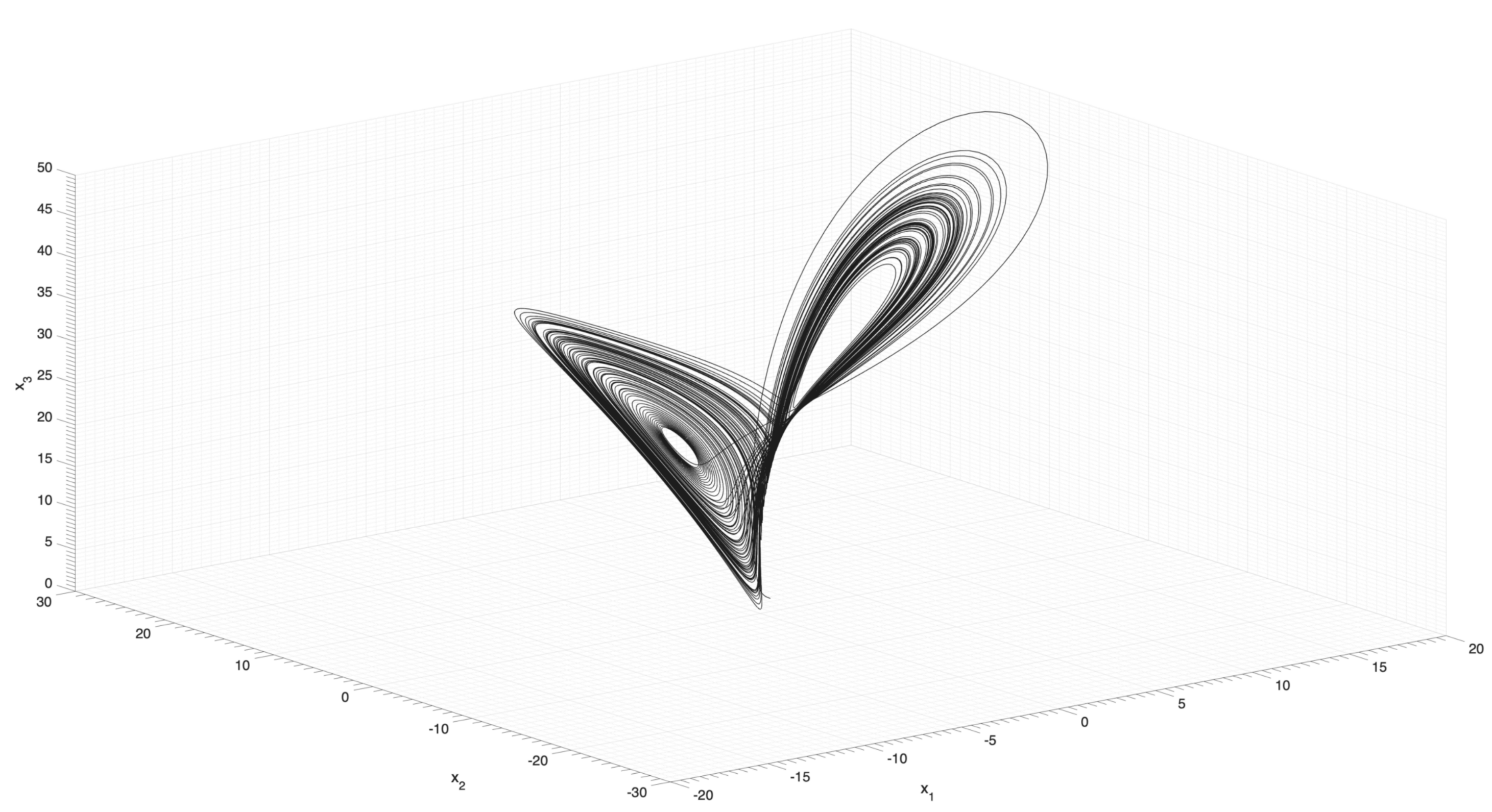}
  \caption{The Lorenz strange attractor}
  \label{lorat}
\end{figure}

\begin{figure}
  \centering
  % include second image
  \includegraphics[width=1\linewidth]{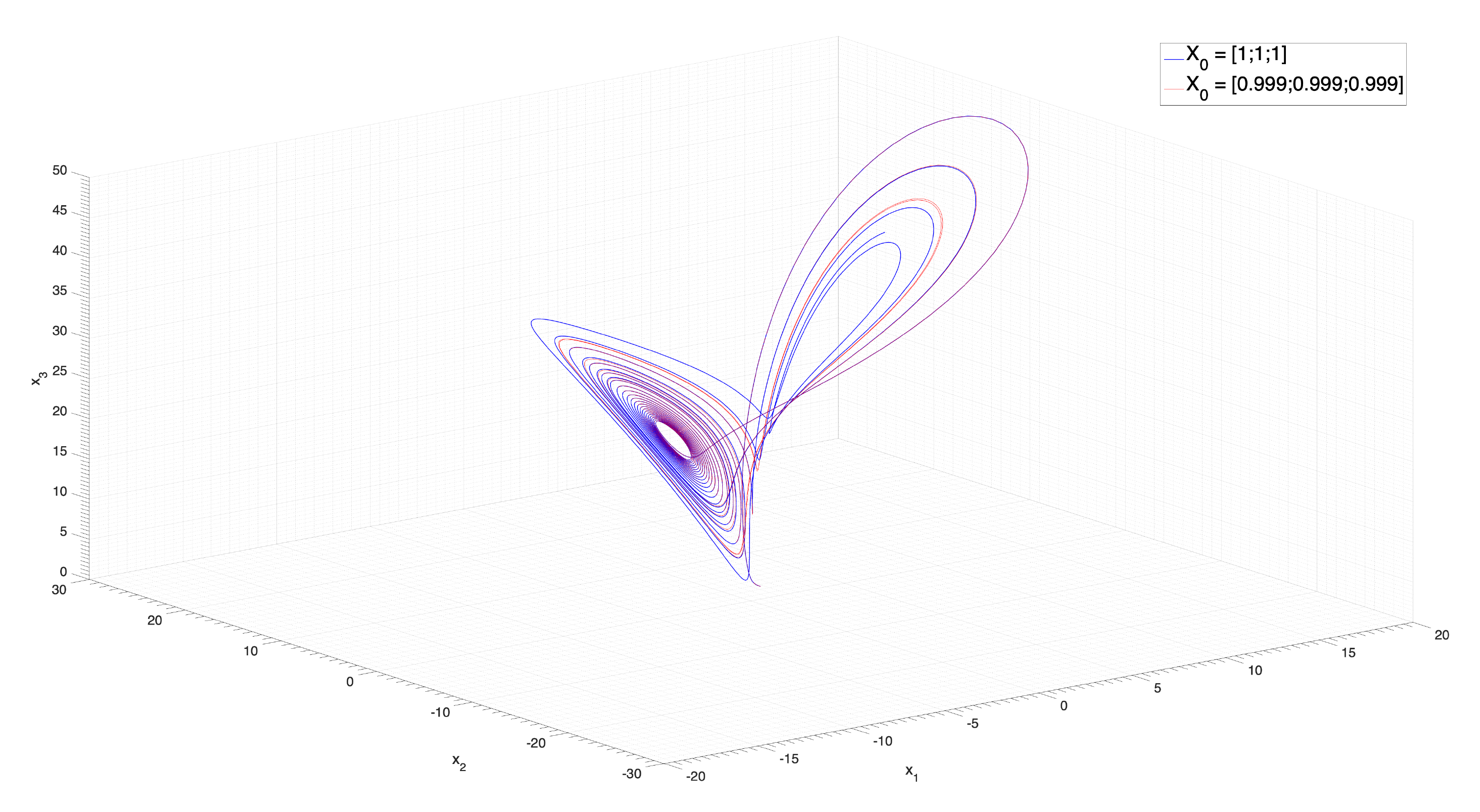}
  \caption{The sensitivity to initial conditions}
  \label{loratinit}
\end{figure}

\begin{figure}
  \centering
  % include first image
  \includegraphics[width=1\linewidth]{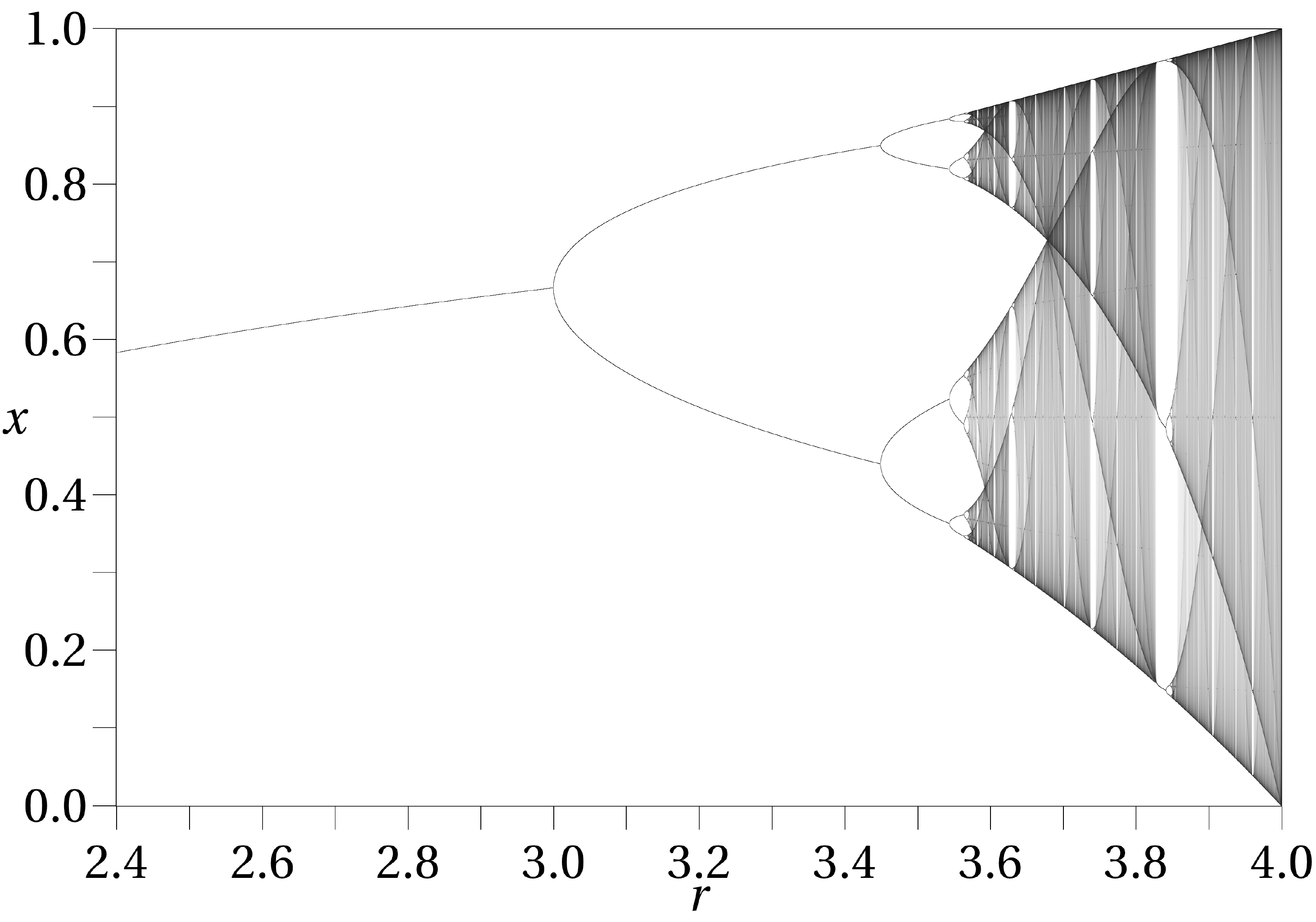}
  \caption{The Bifurcation diagram of the logistic map}
  \label{bif}
\end{figure}

\begin{figure}
  \centering
  % include second image
  \includegraphics[width=1\linewidth]{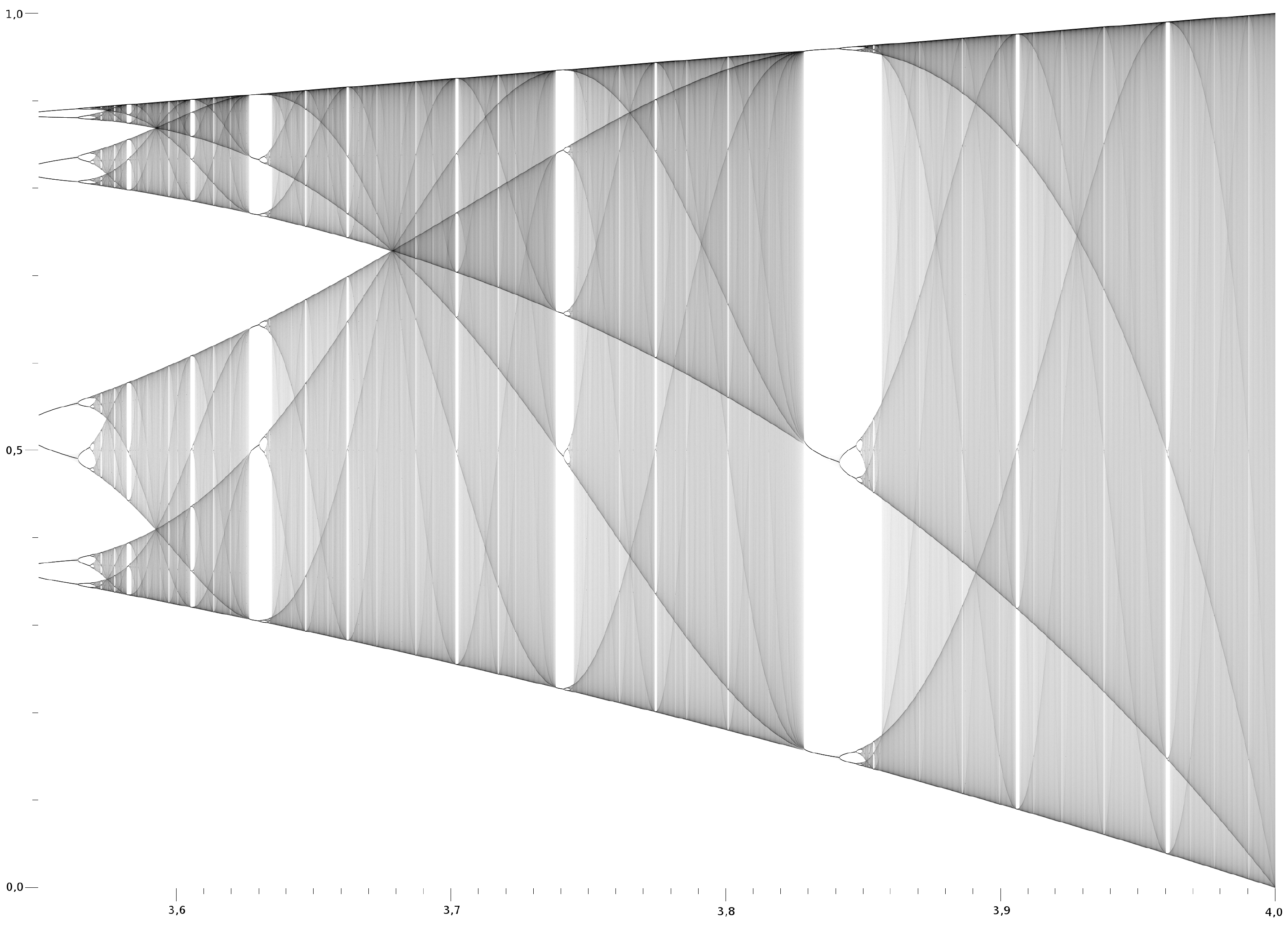}
  \caption{The fractal structure of figure \ref{bif}}
  \label{fractal}
\end{figure}

\begin{figure}
    \centering
    \includegraphics[width=0.45\linewidth]{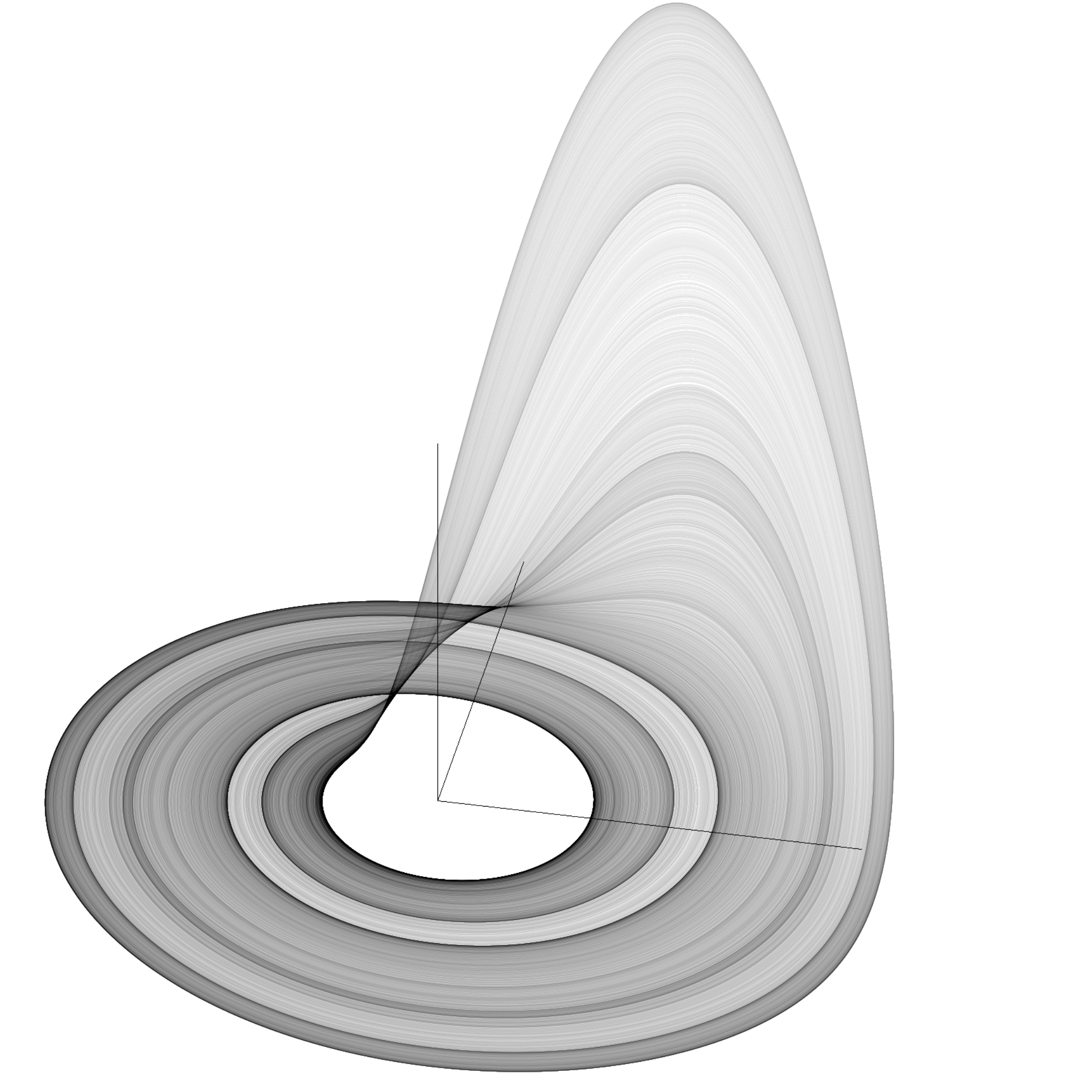}
    \caption{The Rossler attractor}
    \label{rossler}
\end{figure}

Chaos theory is the science that studies deterministic dynamical systems having a high sensitivity to initial conditions. A dynamical system $\dot{x} = f(x)$ has high sensitivity to initial conditions if a tiny change in the initial conditions causes a large change in the trajectory of the solution. The sensitivity to initial conditions of chaotic systems is known to the public as the butterfly effect, in reference to the famous conference by meteorologist Edward Lorenz titled: "Predictability: Does the Flap of a Butterfly's Wings in Brazil Set off a Tornado in Texas?" \cite{Lorenz}. The objective of this title is to support the fact that in order to know the state of a chaotic system such as the atmosphere, all the details on the initial conditions, however small they may be, are essential. Obviously, a butterfly wing flap generally cannot cause a tornado. The fact remains that its effect is immense, as is the effect of any initial condition on a chaotic system. The only reason why a butterfly wing flap generally cannot cause a tornado is that there is an infinity of other effects which help to "regulate" the atmosphere. Lorenz's simplified model of the atmosphere is given by \refeq{lorenzequation}. 
\begin{equation} \label{lorenzequation}
\left\lbrace
\begin{aligned}
&\dot{x}_{1} = \sigma (x_{2}-x_{1}) \\
&\dot{x}_{2} = x_{1}(\rho - x_{3}) - x_{2} \\
&\dot{x}_{3} = x_{1}x_{2} - \beta x_{3}
\end{aligned}
\right.
\end{equation}
Lorenz used : $\sigma = 10$, $\beta = 8/3$ and $\rho = 28$. But for nearby values of these constants, the system will still remain chaotic. For close initial conditions, the trajectories of the system remain close at the beginning and then become different as it is shown in figure \ref{loratinit}. During the simulation, we can see that a certain shape appears (figure \ref{lorat}). This shape is called the Lorenz strange attractor, it is the set of values that can take the vector $[x_{1},x_{2},x_{3}]^T$. The idea behind this strange attractor is that even if the trajectories are very different for different initial conditions, they still remain in a certain bounded set which is the attractor.
\\ \\ The trajectory of a chaotic system for a given initial condition is called an orbit. For continuous time chaotic systems like the Lorenz system, the orbit is a continuous curve. But there are also discrete-time chaotic systems of the form $x_{n+1} = f(x_{n})$ for which the orbit is a set of discrete points. As an example of such systems: The logistic map. It is an idealized model of population growth given by \refeq{logisticmap}.
\begin{equation} \label{logisticmap}
x_{n+1} = rx_{n}(1-x_{n})
\end{equation}
The parameter that tells us if the system is chaotic or not is the value $r$. For example, if we take $r = 2$, the system will converge to the value $0.5$ for any initial condition, so the system is not chaotic. For $r=3.1$, the system oscillates between two values: $0.76$ and $0.56$, it is called a periodic orbit. For $r=3.57$, we have no longer oscillations of finite periods (because we have oscillations between an infinite number of values). It is the beginning of Chaos. We can do a graph of the different values that can take the system for a large time with respect to $r$. This graph is called a bifurcation diagram and is shown in figure \ref{bif}. A bifurcation is a change in the period length of the orbits. We can still see some places beyond $r=3.57$ where there is no chaos, if we zoom in as in figure \ref{fractal}, we can remark a fractal structure.
\\ \\ We need sometimes to add an external input $m$ without losing the chaotic behavior of the chaotic system. A system with external input is a system of the form $\dot{x} = f(x,m)$ where $m$ is the external input. One of them is the modified Rossler's chaotic system which is presented in \cite{dimassi} and is given by \refeq{lorenzexternal}.
\begin{equation} \label{lorenzexternal}
\left\lbrace
\begin{aligned}
&\dot{x}_{1} = - (x_{2}+x_{3}) \\
&\dot{x}_{2} = x_{1} + ax_{2} \\
&\dot{x}_{3} = b + x_{3}(x_{1}-c) + mx_{3}
\end{aligned}
\right.
\end{equation}
For $a=0.398$, $b=2$ and $c=4$, this system is chaotic. The attractor of this system has the form shown in figure \ref{rossler}.
\subsection{The concept of chaos synchronization}
\begin{figure}
    \centering
    \includegraphics[width=0.65\linewidth]{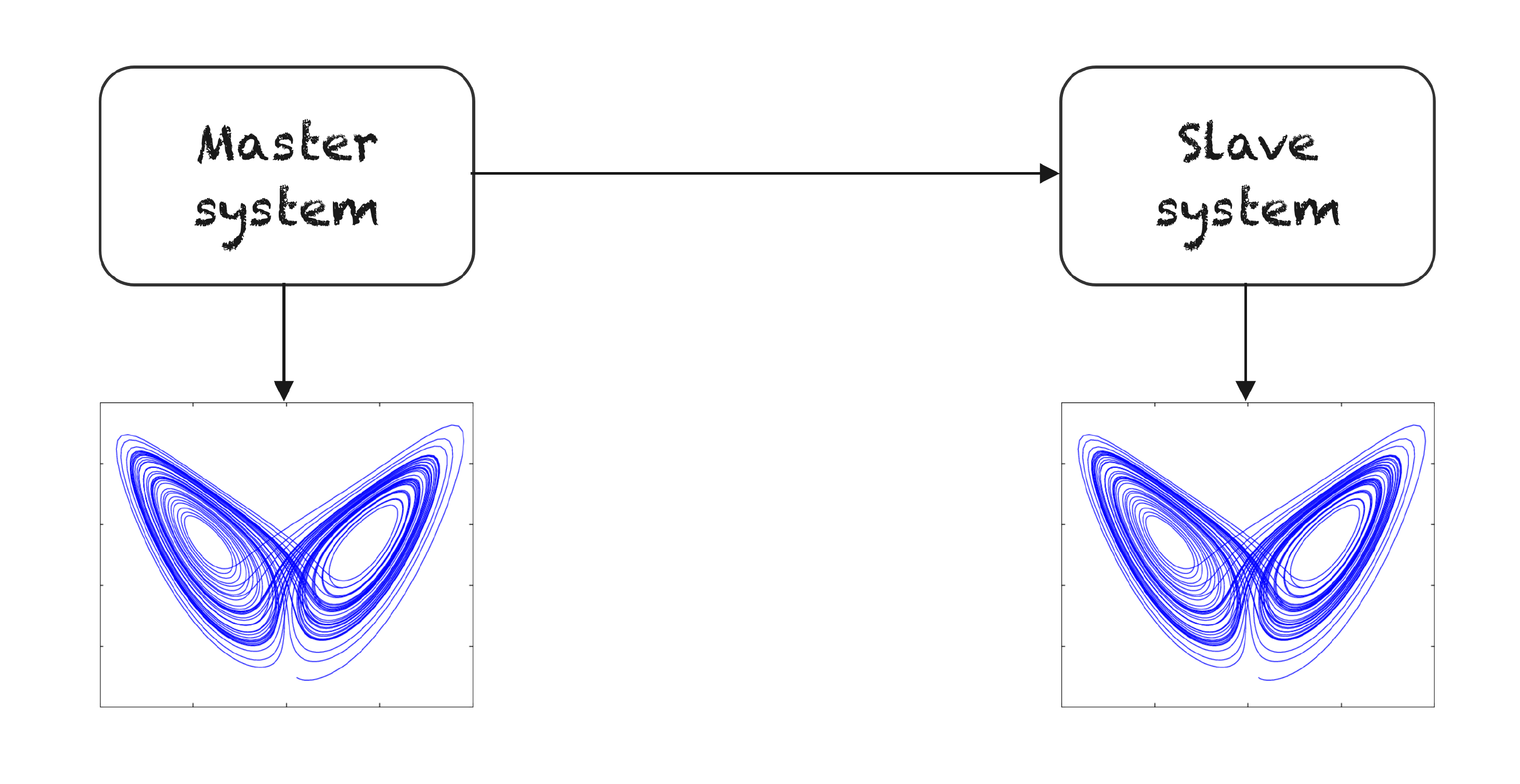}
    \caption{The principle of chaos synchronization}
    \label{master}
\end{figure}

 Chaotic systems can be implemented in hardware (analog form) or in software (digital form). For the digital case, there is a phenomenon called dynamical degradation, which means that the chaotic behavior of the system will become non-ideal. In this paper, I focus exclusively on the analog form of chaotic systems. These systems are designed as electrical circuits. There is however a very important problem: the synchronization of chaotic systems. In chaotic cryptography, we need two versions of the same chaotic signal, one at the emitter and the other at the receiver. We can try to design the same chaotic circuit at the emitter and at the receiver but the parameters will of course never be exactly the same in the two sides of the communication system because of noise and uncertainties. The problem is that chaotic systems are very sensitive to initial conditions, so it is practically impossible to generate the same chaotic signal by this technique. Chaos synchronization is the science that studies how to generate the same chaotic signal as a given chaotic circuit. It really started in 1990 with the works of Pecora and Caroll \cite{pecora}. The idea is that there is a master chaotic system that will send a signal to another dynamical system (called the slave system) for synchronization so that the slave system generates the chaotic trajectory of the master chaotic system as it is shown in figure \ref{master}.
There are a lot of chaos synchronization techniques. We have for example the chaos synchronization based on state-feedback control, the chaos synchronization based on Backstepping, and the chaos synchronization based on non-linear observers, which is the subject of this paper. The idea behind the chaos synchronization based on non-linear observers is that at the slave system is an observer that will try to reconstruct the master system's signals.
\subsection{The design of chaotic cryptosystems}
\begin{figure}
    \centering
    \includegraphics[width=1\linewidth]{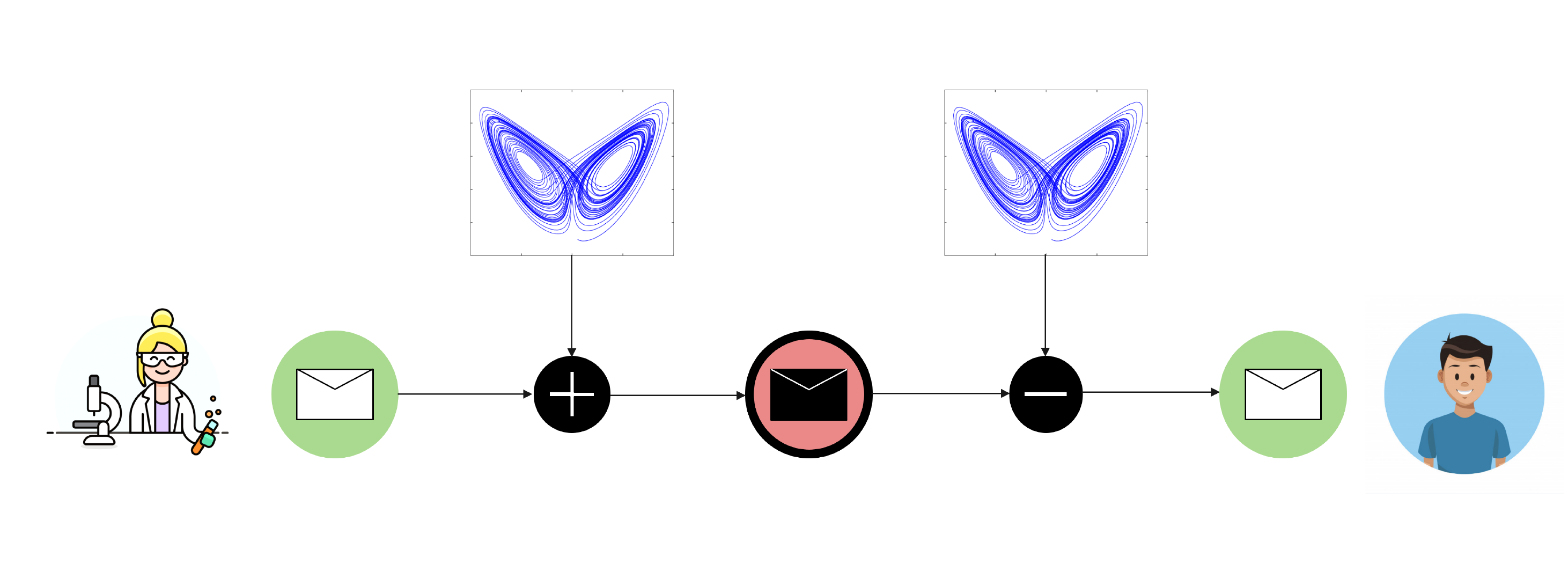}
    \caption{Chaotic masking}
    \label{masking}
\end{figure}

\begin{figure}
    \centering
    \includegraphics[width=1\linewidth]{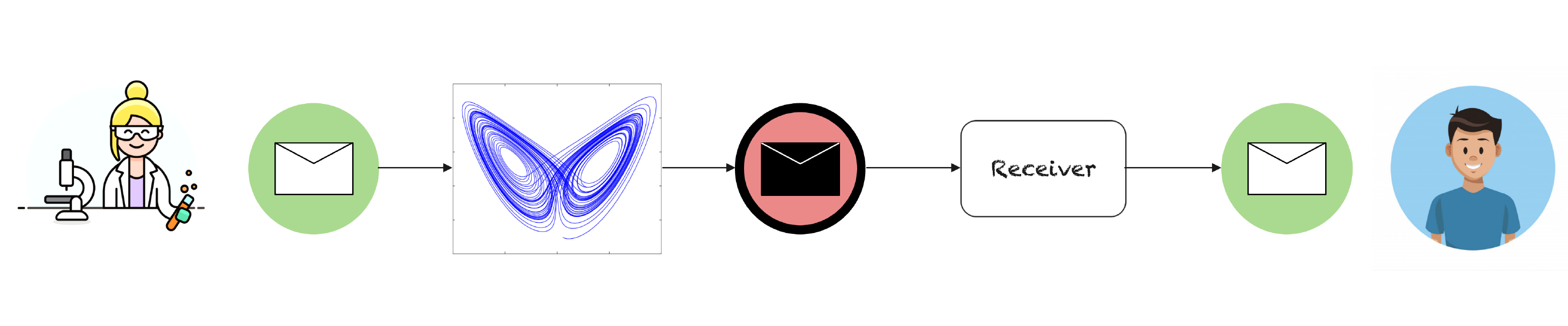}
    \caption{Chaotic modulation}
    \label{modulation}
\end{figure}

I will present now two designs that are widely used in the literature \cite{chaos1,chaos2,chaos3,chaos4}. These designs are chaotic masking and chaotic modulation.  
\\ \\ - \textbf{Chaotic masking:}
The structure of chaotic masking is shown in figure \ref{masking}. The idea is to add to the plaintext the chaotic signal, and then subtract it at the receiver side. It can be done in analog form by adding the chaotic signal to the binary signal, or in the digital form by first converting the chaotic signal into a sequence of bits and then XORing the result with the digital plaintext.
\\ - \textbf{Chaotic modulation:}
The structure of chaotic modulation is shown in figure \ref{modulation}. The idea is to use the plaintext as an external input to the emitter's chaotic system. Then, at the receiver we reconstruct this external input. 

\section{Nonlinear Observers.}
As we have seen before, we need to synchronize chaos. In this paper, this synchronization is achieved using non-linear observers.
\\ \\ The goal of an observer is in general to find the state $x$ (or part of it) of a dynamical system using some knowledge about the output and the input of the system. A dynamical system is described by the general state equation \refeq{dynamics}.
\begin{equation} \label{dynamics}
\left\lbrace
\begin{aligned}
&\dot{x}= f(x,m) \\
&y= h(x)
\end{aligned}
\right.
\end{equation}
where $x\in \mathbf{R}^n$, $y\in \mathbf{R}^p$ and $m\in \mathbf{R}^q$. The problem of a state observer is to determine $x$ (or part of it) knowing $m$ and $y$. Sometimes $m$ is not known and the goal of the observer is to estimate $x$ and $m$ using $y$. Before we can design a state observer, we need to know if it is possible to observe the state vector $x$. This gives rise to the notion of observability of dynamical systems. The question is the following: Given a dynamical system represented by the functions $f$ and $h$, and knowing $y$ and $u$, can we reconstruct $x$ ?
\subsection{The observability problem}
The notion of observability depends on the output vector $y$, the input vector $m$ and the structure of the dynamical system (the functions $f$ and $h$). Let us first analyze the observability with respect to $y$.
\\ \\ For a given input signal $m$, let us consider the function \refeq{phi}. 
\begin{align} \label{phi}
\Phi_{m} : \ \ &\mathbf{I} \ \ \to \Gamma \\
&x_{0} \ \mapsto \Phi_{m}(x_{0}) = h(\lbrace x_{m}(t,x_{0}) \rbrace ) \nonumber
\end{align}
where $I\subset R^n$ is the set of all initial conditions of the dynamical system, $\Gamma $ is the set of all the curves on $\mathbf{R}^p$ and $\lbrace x_{m}(t,x_{0}) \rbrace$ is a solution for the input $m$ and given the initial condition $x_{0}$\footnote{The function $\Phi_{m}$ is well defined iff the Cauchy-Lipschitz theorem holds.}. This function allows us to define the observability given a certain input signal $m$. All the following definitions concerning observability are reformulations I did of the definitions of \cite{besancon} using the function $\Phi_{m}$. \\ \\ \textbf{Definition 2:} \textit{Observability \\
The system \refeq{dynamics}, under the application of an input signal $m$, is said to be observable if $\Phi_{m}$ is injective. In other words, there is no ambiguity in the state trajectory given a certain output and a certain input.} 
\\ \\ \textbf{Definition 3:} \textit{Weak observability \\
The system \refeq{dynamics}, under the application of an input signal $m$, is said to be weakly observable if for every $x_{0}\in \mathbf{I}$ there is a neighborhood $V$ of $x_{0}$ such that the restriction of $\Phi_{m} $ to $V$ is injective.}
\\ \\ Sometimes, observability will appear only after a certain period of time. It is the case for example of two state curves that have the same output during an interval of time $[t_{0},t_{f}]$ and whose corresponding outputs are not necessarily equal after $t_{f}$. This gives rise to the notion of local weak observability: \\ \\
\textbf{Definition 4:} \textit{Local weak observability \\
The system \refeq{dynamics}, under the application of an input signal $m$, is said to be locally weakly observable if for every $x_{0}\in \mathbf{I}$ there is a neighborhood $V$ of $x_{0}$ such that for any neighborhood $W$ of $x_{0}$ contained in $V$, the restriction of $\Phi_{m}$ to $W$ is injective when considering time intervals for which trajectories remain in $V$.} 
\\ \\ In practice, we have a formula to check if the system is observable or not. This formula is known as the observability rank condition. 
\\ \\ \textbf{Definition 5 \cite{besancon}:} \textit{The observability rank condition \\
The system \refeq{dynamics}, under the application of an input signal $m$, is said to satisfy the observability rank condition if \refeq{rank} holds.
\begin{equation} \label{rank}
rank\lbrace [\frac{\partial h(x)}{\partial x} \ \ \frac{\partial L_{f} h(x)}{\partial x} \ \ \frac{\partial L^2_{f} h(x)}{\partial x} \ \ ... \ \ \frac{\partial L^{n-1}_{f} h(x)}{\partial x}]^{T} \rbrace = n
\end{equation}
where \refeq{lie},
\begin{equation} \label{lie}
L^i_{f}h(x) = L_{f}(L^{i-1}_{f}h(x))
\end{equation}
and $L_{f}h(x)$ is the Lie derivative of $h$ along $f$ defined by \refeq{lieder}.}
\begin{equation} \label{lieder}
L_{f} h(x) = \frac{\partial h(x)}{\partial x}f(x,m)
\end{equation}
\\ For linear time invariant (LTI) systems ($f(x,m) = Ax + Bm$ and $h(x)= Cx$), we have \refeq{lti}. 
\begin{equation} \label{lti}
h(x) = Cx \Rightarrow \frac{\partial h(x)}{\partial x} = C \Rightarrow \frac{\partial L_{f} h(x)}{\partial x} = CA \Rightarrow ... \Rightarrow \frac{\partial L^{n-1}_{f}h(x)}{\partial x} = CA^{n-1}
\end{equation}
which gives us the observability rank condition for LTI systems \refeq{ltieq}.
\begin{equation} \label{ltieq}
rank\lbrace [C \ \ CA \ \ CA^2 \ \ ... \ \ CA^{n-1}]^{T} \rbrace = n
\end{equation}
The relation between observability and the observability rank condition is given by the following theorem:
\\ \\ \\ \\ \textbf{Theorem \cite{besancon}:} \textit{ \\
If the system \refeq{dynamics}, under the application of an input signal $m$, satisfies the observability rank condition, then it is locally weakly observable. Conversely, if the system \refeq{dynamics} is locally weakly observable, then it satisfies the observability rank condition in an open dense subset of $\mathbf{R}^n$.} \\ \\
We will now focus our attention on the observability with respect to the input vector $m$. For a certain input signal $m$, we can study the observability of the system with respect to the output. This observability can of course be affected by the input, which gives rise to the following definition: \\ \\
\textbf{Definition 6 \cite{besancon}:} \textit{Uniformly observable systems \\
A system is uniformly observable (UO) if it is observable for any input signal $m$. If the system is (UO) only in an interval of time $[0,t]$, we say that it is locally uniformly observable. } \\ \\ For a system that is (UO), the observability does not depend on the input.
\subsection{Classical observers}
Once we know if the dynamical system is observable, we can design observers. I will start with classical observers which are of two types: Linear observers and non-linear observers. Then, I will describe the more advanced observers. The part concerning advanced observers is strongly inspired by the state of the art presented in \cite{dimassi}.
\subsubsection{Linear Observers}
$\bullet $ \textit{\underline{Kalman filter for stochastic LTI systems :}} \\ \\
Let us consider the stochastic LTI system given by \refeq{kalman1}.
\begin{equation} \label{kalman1}
\left\lbrace
\begin{aligned}
&\dot{x} = Ax + Bm + w\\
&y = Cx+v
\end{aligned}
\right.
\end{equation}
where $w$ is the process noise and $v$ is the measurement noise (these noises are assumed to be Gaussian). If \refeq{kalman1} is observable, then there exists an observer of the form \refeq{observerkalman}.
\begin{equation} \label{observerkalman}
\dot{\hat{x}} = A\hat{x} + Bm - L(C\hat{x}-y)
\end{equation}
where $L = PM^{T}R^{-1}$, such that $P$ satisfies the Algebraic Riccati equation (ARE) given by \refeq{ARE}.
\begin{equation} \label{ARE}
AP + PA^T + Q - PM^TR^{-1}MP = 0
\end{equation}
where $R$ and $Q$ are the covariance matrices of $v$ and $w$ respectively. 
\\ \\ \\ $\bullet $ \textit{\underline{Luenberger observer for deterministic LTI systems:}} \\ \\
Let us consider the deterministic LTI system given by \refeq{luenberger}. 
\begin{equation} \label{luenberger}
\left\lbrace
\begin{aligned}
&\dot{x} = Ax + Bm\\
&y = Cx
\end{aligned}
\right.
\end{equation}
If \refeq{luenberger} is observable, then there exists an observer of the form \refeq{luenbergerobserver}.
\begin{equation} \label{luenbergerobserver}
\dot{\hat{x}} = A\hat{x} + Bm - L(C\hat{x}-y)
\end{equation}
where $L$ is such that $(A-LC)$ is stable.
\subsubsection{Non-linear Observers}~\\ \\
$\bullet $ \textit{\underline{Extended Kalman filter:} }\\ \\
In this method we linearize \refeq{dynamics} in the neighborhood of our operating point, and then we apply the Kalman filter by considering the non-linearities as noises. 
\\ \\ $\bullet $ \textit{\underline{Thau's observer:} }\\ \\
Let us consider the system given by \refeq{thau}.
\begin{equation} \label{thau}
\left\lbrace
\begin{aligned}
&\dot{x} = Ax + g(t,m,y) + f(x)\\
&y = Cx
\end{aligned}
\right.
\end{equation}
The Thau's observer for \refeq{thau} is given by \refeq{thauobserver}.
\begin{equation} \label{thauobserver}
\dot{\hat{x}} = A\hat{x} + g(t,m,y) + f(\hat{x}) - L(C\hat{x}-y) 
\end{equation}
This observer converges if \refeq{thau} is observable, $f$ is globally Lipschitz with a Lipschitz constant $\gamma$ and if $L$ satisfies an equation of the form \refeq{L}.
\begin{equation} \label{L}
(A-LC)^{T}P+P(A-LC)=-Q
\end{equation}
where $P$ and $Q$ are positive definite matrices that satisfy the inequality \refeq{in}.
\begin{equation} \label{in}
\gamma < \frac{\lambda_{min}(Q)}{2\lambda_{max}(P)}
\end{equation}
\\ $\bullet $ \textit{\underline{High gain observer (Ragahvan's method):} }\\ \\
The High gain observer is for systems of the form \refeq{thau}. Let us assume that the system is observable and that $f$ is globally Lipschitz with a Lipschitz constant $\gamma$. The high gain observer has the same form as Thau's observer except that the determination of $L$ is different. To determine $L$, we use the Ragahvan's method: \\ \\
1- Set $\epsilon > 0$. \\
2- Solve the Riccati equation for $P$ given by \refeq{ric}.
\begin{equation} \label{ric}
AP + PA^T + P(\gamma ^{2} I - \frac{C^TC}{\epsilon } )P + I(\epsilon + 1) = 0
\end{equation}
3- Check if $P$ is symmetric and positive definite \\
(i) If yes, set $L=\frac{PC^T}{2\epsilon}$. \\
(ii) If no, set $\epsilon = \frac{\epsilon }{2}$ and repeat.
\subsection{Sliding mode observers}
Sliding mode observers are very efficient for linear systems with unknown uncertainties. These systems are given by \refeq{sliding}.
\begin{equation} \label{sliding}
\left\lbrace
\begin{aligned}
&\dot{x}= Ax + Bm + Df(x,m,t) \\
&y= Cx
\end{aligned}
\right.
\end{equation}
where $D$ is a matrix of the appropriate dimension and $f$ is the unknown uncertainty. 
\\ \\ Let us assume that $f$ is bounded by some scalar $\rho $, i.e $\ ||f(x,m,t)|| \leq \rho, \ \ \forall x\in \mathbf{R}^n, \forall m\in \mathbf{R}^q, \forall t\geq 0$. There are three types of sliding mode observers which are widely used in practice: The Walcott-Zak sliding mode observer, the Edwards-Spurgeon sliding mode observer and the Higher order sliding mode observers. I will present now each one of them. 
\subsubsection{The Walcott-Zak sliding mode observer}~\\ \\
For the Walcott-Zak sliding mode observer \cite{56}, we add a structural condition on the unknown uncertainty. Namely, we assume that there are two positive definite matrices $P$ and $Q$ and two matrices of appropriate dimensions $L$ and $F$ such that \refeq{WZ}. 
\begin{equation} \label{WZ}
\left\lbrace
\begin{aligned}
&(A-LC)^TP + P(A-LC) = -Q \\
&PD = C^TF^T
\end{aligned}
\right.
\end{equation}
We also assume that the pair $(A,C)$ is observable. The Walcott-Zak observer is given by \refeq{WZo}.
\begin{equation} \label{WZo}
\begin{aligned}
\dot{\hat{x}} = A\hat{x} + Bm - L(C\hat{x}-y)+ \mu
\end{aligned}
\end{equation}
where $\mu $ is a discontinuous function defined by \refeq{mu}.
\begin{equation} \label{mu}
\mu =
\left\lbrace
\begin{aligned}
&-\rho \frac{P^{-1}C^TF^TFCe}{||FCe||}, \ \ &if \ \ FCe\neq 0 \\
&\ \ \ \ \ \ \ \ \ \ \ \ \ \ \ \ \ 0, \ \ &if \ \ FCe=0
\end{aligned}
\right.
\end{equation}
 For details about the convergence of this observer, you can refer to \cite{57}. The principal disadvantage of this technique is that the discontinuity of $\mu$ gives rise to high-frequency oscillations. The apparition of these high-frequency oscillations is called the chattering effect. 
\subsubsection{The Edwards-Spurgeon sliding mode observer}~\\ \\
For Edwards-Spurgeon sliding mode observer \cite{58}, we define the sliding surface $S := \lbrace e\in \mathbf{R}^n:Ce = 0\rbrace $ where $e=x-\hat{x}$. Let us assume that $rank(CD)=rank(D)$ and that the invariant zeroes of $(A,D,C)$ are in $\mathbf{C}_{-}$. It was proved in \cite{58} that under these assumptions there exist a non-singular change of coordinates $x\mapsto Tx$ that transforms \refeq{sliding} into \refeq{tx}.
\begin{equation} \label{tx}
\left\lbrace
\begin{aligned}
&\dot{x}_{1} = A_{11}x_{1} + A_{12}x_{2} + B_{1}m \\
&\dot{x}_{2} = A_{21}x_{1} + A_{22}x_{2} + B_{2}m + D_{2}f(x,m,t) \\
&y = x_{2}
\end{aligned}
\right.
\end{equation}
where $x_{1}\in \mathbf{R}^{n-p}$, $x_{2}\in \mathbf{R}^p$ and $A_{11}$ is Hurwitz. \\ \\
The Edwards-Spurgeon sliding mode observer is given by \refeq{ES}.
\begin{equation} \label{ES}
\begin{aligned}
&\dot{\hat{x}}_{1} = A_{11}\hat{x}_{1} + A_{12}\hat{x}_{2} + B_{1}m \\
&\dot{\hat{x}}_{2} = A_{21}\hat{x}_{1} + A_{22}\hat{x}_{2} + B_{2}m - (A_{22}-A_{22}^s)e_{y} + \nu \\
&\hat{y} = \hat{x}_{2}
\end{aligned}
\end{equation}
where $A_{22}$ is Hurwitz, $e_{y} = y-\hat{y}$ and $\nu$ is a discontinuous function given by \refeq{nu}.
\begin{equation} \label{nu}
\nu =
\left\lbrace
\begin{aligned}
&-\rho ||D_{2}|| \frac{P_{2}e_{y}}{||P_{2}e_{y}||}, \ \ &if \ \ e_{y}\neq 0 \\
&\ \ \ \ \ \ \ \ \ \ \ \ \ \ \ \ \ 0, \ \ &if \ \ e_{y}=0
\end{aligned}
\right.
\end{equation}
where $P_{2}$ is a Lyapunov matrix for $A_{22}^s$. The state estimate is given by \refeq{state}.
\begin{equation} \label{state}
\hat{x} = T^{-1}[\hat{x}_{1}, \hat{x}_{2}]^T
\end{equation}
One problem with Edwards-Spurgeon sliding mode observer is that the hypotheses are not always verified.
\subsubsection{Higher-order sliding mode observers}~\\ \\
Higher-order sliding mode observers \cite{59,60,61,62} are for dynamical system in the triangular form \refeq{dynamicaltri}.
\begin{equation} \label{dynamicaltri}
\left\lbrace
\begin{aligned}
&\dot{x} = A_{n} x + H_{n}V_{n}(x,w) \\
&y = C_{n}x
\end{aligned}
\right.
\end{equation}
where $w\in \mathbf{R}$ is an unknown input and 
\begin{equation}
A_{n} = \begin{bmatrix} 0 & 1 & ... & 0\\
                        0 & 0 & 1 & ... \\
                        ... & ... & ... & 1\\
                        0 & ... & ... & 0\end{bmatrix}, \ \ H_{n} = \begin{bmatrix} 
                        0 & ... & 1 \end{bmatrix}^T, \ \ C_{n} = \begin{bmatrix}
                        1 & 0 & ... & 0 \end{bmatrix} \nonumber
\end{equation}
Let us assume that the state is uniformly bounded, i.e $\exists (d_{1},...,d_{n})\in \mathbf{R}^n , \ s.t \ \forall t >0, \forall i\in \lbrace 1,2,...,n \rbrace : \ |x_{i}(t)| < d_{i} $.  Let us also assume that $w$ and its derivative are bounded. 
\\ \\ There are several types of Higher-order sliding mode observers. For example, the Higher-order sliding mode observer presented in \cite{59}, which has the form \refeq{HO} is a second order sliding mode observer.  
\begin{equation} \label{HO}
\begin{aligned}
&\dot{\hat{x}}_{1} = z_{1} + \lambda_{1}\sqrt{|x_{1}-\hat{x}_{1}|} sign(x_{1}-\hat{x}_{1}) \\
&\dot{z}_{1} = \alpha_{1}sign(x_{1}-\hat{x}_{1}) \\
&\dot{\hat{x}}_{2} = z_{2} + \lambda_{2}\sqrt{|z_{1}-\hat{x}_{2}|} sign(z_{1}-\hat{x}_{2}) \\
&\dot{z}_{2} = \alpha_{2}sign(z_{1}-\hat{x}_{2}) \\
&... \\
&\dot{\hat{x}}_{n-1} = z_{n-1} + \lambda_{n-1}\sqrt{|z_{n-2}-\hat{x}_{n-1}|} sign(z_{n-2}-\hat{x}_{n-1}) \\
&\dot{z}_{n-1} = \alpha_{n-1}sign(z_{n-2}-\hat{x}_{n-1}) \\
&\dot{\hat{x}}_{n} = z_{n} + \lambda_{n}\sqrt{|z_{n-1}-\hat{x}_{n}|} sign(z_{n-1}-\hat{x}_{n}) \\
&\dot{z}_{n} = \alpha_{n}sign(z_{n-1}-\hat{x}_{n})
\end{aligned}
\end{equation}
where $\lambda_{i}$ and $\alpha_{i}$ are the observer gains. They are positive scalars that we need to define. There are other higher order sliding mode observers like third order sliding mode observers \cite{Third} and fourth order sliding mode observers \cite{Fourth}. The main advantage of Higher order sliding mode observers over classical sliding mode observers is that the chattering effect is reduced. It does not mean that there is no chattering effect in Higher order sliding mode observers, it is just reduced. If we want to eliminate it almost completely we can replace the discontinuous functions as the sign function with fuzzy inference systems \cite{fuzzy1,fuzzy2}.
\subsection{Unknown inputs observers}
The goal of an unknown inputs observer is to estimate the state of a dynamical system without having knowledge about part of (or all) the input. There are a lot of techniques of design of unknown inputs observers. One of them is to separate the state vector into two parts: one part which is influenced by the unknown inputs and the other which is not. It is possible to design with this method a reduced order observer \cite{63}. Other methods are called algebraic methods of design and they are based on the resolution of matrix linear equations \cite{64}. In the algebraic methods of design, we consider linear systems of the form \refeq{UI}.
\begin{equation} \label{UI}
\left\lbrace
\begin{aligned}
&\dot{x} = Ax + Bu + Fw\\
&y = Cx
\end{aligned}
\right.
\end{equation}
where $u$ is the known input vector and $w$ the unknown input vector. Let us assume that $F$ is full rank and that $(A,C)$ is observable. The full order unknown input observer presented in \cite{64,65} is given by \refeq{FO}.
\begin{equation} \label{FO}
\begin{aligned}
&\dot{z} = Nz + Gu + Ly\\
&\hat{x} = z-Ey
\end{aligned}
\end{equation}
where $N$, $G$, $L$ and $E$ are matrices that we need to determine such that the observer converges. The dynamics of the observer error $e=x-\hat{x}$ is given by \refeq{errordyn}.
\begin{equation} \label{errordyn}
\dot{e} = Ne + (PB - G)u + (PA - NP - LC)x
\end{equation}
where $P = I + EC$. The error $e$ converges asymptotically to zero if and only if $N$ is stable, $P = I + EC$, $LC =PA - NP$, $G = PB$ and $PF = 0$. It is a set of matrix linear equations. The sufficient and necessary conditions for the existence of a solutions to these equations are given by \refeq{condi}.  
\begin{equation} \label{condi}
\begin{aligned}
&(i) \ \ rank(CF) = rank(F) \\
&(ii) \ \ rank\begin{bmatrix} sP-PA\\ C \end{bmatrix} = n, \ \ \forall s\in \mathbf{C}, \ \ Re(s) \geq 0
\end{aligned}
\end{equation}
 In \cite{66}, this result is extended to the case where the unknown inputs affect the state and affect also the output as in \refeq{UI2}. 
\begin{equation} \label{UI2}
\left\lbrace
\begin{aligned}
&\dot{x} = Ax + Bu + F_{1}w\\
&y = Cx + F_{2}w
\end{aligned}
\right.
\end{equation}
Let us assume that \refeq{UI2} satisfies a structural constraint of the form \refeq{structural}.
\begin{equation} \label{structural}
\begin{aligned}
&rank\begin{bmatrix} CF_{1} & F_{2}\\ F_{2} & 0\end{bmatrix} = rank(G) + rank\begin{bmatrix}F_{1}\\ F_{2} \end{bmatrix}\\
&rank\begin{bmatrix}sI-A & -F_{1}\\ C & F_{2}\end{bmatrix} = n + rank\begin{bmatrix} F_{1}\\ F_{2}\end{bmatrix}, \ \ \forall s\in \mathbf{C}, \ \ Re(s) \geq 0
\end{aligned}
\end{equation}
The unknown inputs observer presented in \cite{66} for this system is given by \refeq{UI3}. 
\begin{equation} \label{UI3}
\begin{aligned}
&\dot{z} = Nz + Hu + Jy\\
&\hat{x} = z-Ey
\end{aligned}
\end{equation}
The observer error converges asymptotically to zero if and only if $N$ is stable, $P = I+EC$, $PA-NP-JC=0$, $PF_{1}-NEC-JC=0$, $EF_{2}=0$ and $H=PB$.
\\ \\ The last category of unknown inputs systems that we consider are the singular systems with unknown inputs given by \refeq{S}. 
\begin{equation} \label{S}
\left\lbrace
\begin{aligned}
&E\dot{x} = Ax + Bu + Fw\\
&y = Cx
\end{aligned}
\right.
\end{equation}
The idea is to consider an augmented state vector $\bar{x} = \begin{bmatrix}x\\w\end{bmatrix}$ and the corresponding augmented system \refeq{augmented}. 
\begin{equation} \label{augmented}
\left\lbrace
\begin{aligned}
&\bar{E}\dot{\bar{x}} = \bar{A}\bar{x} + \bar{B}u\\
&y = \bar{C}\bar{x}
\end{aligned}
\right.
\end{equation}
where 
\begin{equation}
\bar{E} = \begin{bmatrix} E & 0\\ 0 & I\end{bmatrix}, \ \ \bar{A} = \begin{bmatrix} A & N\\ 0 & 0\end{bmatrix}, \bar{B} = \begin{bmatrix} B\\ 0\end{bmatrix}, \ \ \bar{C} = \begin{bmatrix} C & 0\end{bmatrix} \nonumber
\end{equation}
An unknown inputs observer for \refeq{augmented} is given in \cite{67,68} by \refeq{augmentedo}.
\begin{equation} \label{augmentedo}
\begin{aligned}
&\dot{z} = Rz + Hu + Ly\\
&\hat{x} = Mz-Ny
\end{aligned}
\end{equation}
If \refeq{augmented} satisfies the structural conditions \refeq{structural2}.
\begin{equation} \label{structural2}
\begin{aligned}
&rank\begin{bmatrix} s\bar{E}-\bar{A}\\\bar{C} \end{bmatrix} = n, \ \ \forall s\in \mathbf{C}\\
&rank\begin{bmatrix}sI-R\\M\end{bmatrix} = n, \ \ \forall s\in \mathbf{C}
\end{aligned}
\end{equation}
then there exists a matrix $K$ of appropriate dimension such that $Kx-\hat{x} \longrightarrow 0$ as $t\longrightarrow +\infty $, $\forall x_{0},z_{0}$ if and only if there exists a matrix $P$ such that $R$ is stable, $P\bar{A}-RP\bar{E} - L\bar{C} = 0$, $K=MP\bar{E}+N\bar{C}$ and $H=P\bar{B}$.
\subsection{Adaptive observers}
The goal of adaptive observers is to estimate the state vector and parameters of the system. Let us consider the class of systems with unknown parameters given by \refeq{UP}. 
\begin{equation} \label{UP}
\left\lbrace
\begin{aligned}
&\dot{x} = f(x,u,t) + g(x,u,t)\theta \\
&y = h(x)
\end{aligned}
\right.
\end{equation}
where $\theta \in \mathbf{R}^q$ is the unknown parameters vector. An adaptive observer for \refeq{UP} which had been proposed in \cite{51}. It is given by \refeq{AD}.
\begin{equation} \label{AD}
\begin{aligned}
&\dot{\hat{x}} = f(y,\hat{z}u,t) + g(y,\hat{z},u,t)\hat{\theta } + k(h(\hat{x}-y,t) \\
&\hat{x} = [\hat{y},\hat{z}]^T
\end{aligned}
\end{equation}
where $\hat{\theta }$ is updated using the adaptation law \refeq{ADD}.
\begin{equation} \label{ADD}
\dot{\hat{\theta }} = -\Lambda \phi ^T(\hat{y}-y,y,\hat{z},u,t)
\end{equation}
where $\Lambda = \Lambda^T > 0$. To ensure the convergence of $\hat{\theta }$, the function $g$ must satisfy the persistent excitation condition \cite{51}: $\exists T, k_{1}, k_{2} > 0 $ s.t $\forall t \geq 0$ we have \refeq{Pe}. 
\begin{equation} \label{Pe}
k_{1}I_{q} \geq \int_{t}^{t+T} g(y(\tau),\hat{z}(\tau),u(\tau),\tau)g^T(y(\tau),\hat{z}(\tau),u(\tau),\tau)d\tau \geq k_{2}I_{q}
\end{equation}
The adaptive observer \refeq{AD} for the system \refeq{UP}, with $\dot{\theta} = 0$, converges asymptotically if there exists a decreasing positive definite function $V(t,e)$, with $e=\hat{x}-x = [\hat{y}-y,\hat{z}-z]^T = [e_{y},e_{z}]^T$, of class $\mathcal{C}^1$, with $|(\frac{\partial V}{\partial e})(t,e)|$ a decreasing function, and a continuous function $k(e_{y},t)$ bounded with respect to $t$ with $k(0,t)=0$, such that $\forall u$, $\forall e$, $\forall y$, $\forall \sigma $, $\forall \alpha >0$, $\forall t \geq 0$, we have \refeq{pfe}.  
\begin{equation} \label{pfe}
\begin{aligned}
&(i) \ \ \dot{V} + \frac{\partial V}{\partial e}[f(y,\sigma,u,t)-f(y,\sigma - e_{z},u,t)+\\ 
&(g(y,\sigma,u,t)-g(y,\sigma-e_{z},u,t))\theta + k(e_{y},t)] \leq -\alpha |e|^2\\
&(ii) \ \ \frac{\partial V}{\partial e} g(y,\sigma,u,t) = \phi(e_{y},y,\sigma,u,t)
\end{aligned}
\end{equation}
and $g$ is globally bounded and $f,g$ are globally Lipschitz with respect to $z$, uniformly with respect to $(u, y, t)$. \\
If in addition to this, the function $g$ satisfies the persistant excitation condition and $\dot{g}$ is bounded, then $||\hat{\theta}-\theta|| \longrightarrow 0$ as $t\longrightarrow + \infty$. \\ \\
A special case of \refeq{UP} is given by \refeq{UP2} \cite{52}. 
\begin{equation} \label{UP2}
\left\lbrace
\begin{aligned}
&\dot{x} = Ax + \psi_{1}(u,x) + B\psi_{2}(u,x)\theta \\
&y = Cx
\end{aligned}
\right.
\end{equation}
where $\psi_{1}$ and $\psi_{2}$ are assumed to be globally Lipshitz with Lipshitz constants $k_{1}$ and $k_{2}$ respectively. We also assume that \refeq{UP2} is of minimum phase and that there exists two positive definite matrices $P$, $Q$ and a matrix $L$ such that \refeq{conditionswaw}.
\begin{equation} \label{conditionswaw}
\begin{aligned}
&P(A - LC) + (A - LC)^TP = -Q\\
&PB = C^T\\
&k_{1} + k_{2}max(\theta) |B| < \frac{\lambda_{min}(Q)}{2\lambda_{max}(P)}
\end{aligned}
\end{equation}
It is possible to construct under these conditions an adaptive observer \cite{52}. 
\\ \\ We can find generalization of adaptive observers for linear MIMO systems in \cite{53}. Other generalizations have been done for non-linear uniformly observable SO (single output) systems \cite{54} and non-linear uniformly observable MIMO systems with non-linear parametrization \cite{55}.
\subsection{ANFIS (Adaptive Neuro-Fuzzy Inference Systems) observers}
An ANFIS observer is a neural network whose training optimizes a fuzzy system. For state estimation using ANFIS, one idea would be to train an ANFIS to predict the states given certain information as the output and the input of the dynamical system. We can also use ANFIS to predict the unknown input of a dynamical system given its state vector (or its output vector or both). In all these cases the principle is the same: we train a neural network to predict some signal given a set of other signals. ANFIS can be used in multiple ways.
\section{Conclusion}
In this survey article I presented an overview of chaos synchronizations using nonlinear observers and its applications in cryptography. I started with a brief overview of classical cryptography. Then, I recalled the basics of chaotic systems and how they could be used in cryptography. I also exposed the problem of synchronization of chaotic systems which is of fundamental importance in chaotic cryptography. I have also recalled the theory of non-linear observers. First, I introduced the notion of observability of a dynamical system, then I presented how to design observers. The observers that have been presented are first of all the classical observers, whether linear or non-linear, then the more advanced observers: Sliding mode observers, Unknown inputs observers, Adaptive observers and ANFIS observers.

\bibliography{bibli}
\bibliographystyle{apalike}

\end{document}